\documentclass[12pt]{iopart}
\usepackage{amssymb}
\usepackage{graphicx}

\def\k{k_{\mbox{\tiny{B}}}}

\def\be{\begin{equation}}
\def\ee{\end{equation}}
\def\ed{\end{document}}
\def\T{T_e}
\def\ba{\begin{eqnarray}}
\def\ea{\end{eqnarray}}
\def\n{\mathbf n}
\def\r{\mathbf r}
\def\v{\mathbf v}
\def\p{\mathbf p}
\def\t{\tilde{t}}

\def\ko{\mbox{\it\tiny{Ko}}}
\def\ze{\mbox{\it\tiny{\tiny{ZL}}}}

\def\h{\hbar\omega}
\def\nt{\tilde{n}}

\begin{document}

\title{Photon plasma-wave interaction via Compton scattering}

\author{G. Erochenkova, C. Chandre}
\address{Centre de Physique Th\'eorique UMR 7332, CNRS -- Aix-Marseille Universit\'e, 13288 Marseille, France}

\begin{abstract}
The Kompaneets theory of photon kinetic evolution due to the Compton effect in the absence of absorption and emission is extended to the case of the Vlasov plasma wave oscillations. Under the assumption that the electron distribution function at equilibrium is perturbed by a solution of the linearised Vlasov equation in the long-wavelength limit, a solution of the Kompaneets kinetic equation for the photon distribution function is found and discussed.
\end{abstract}

\maketitle

\section{Introduction}

The importance of scattering processes resulting from the interaction of electromagnetic radiation (or photons) with charged particles (here electrons) via the Compton effect was noted for the first time by Kompaneets~\cite{KO} in the case of a plasma described by a Maxwellian distribution. An equation for the distribution function of the photons in energy space was derived in this case. In Ref.~\cite{ZL}, Zel'dovitch and Levitch exhibited a singular solution in the case of large occupation numbers, which diverges at a finite time. This singular solution corresponds to the accumulation of photons in the state of zero-energy and was referred to as a Bose condensation of photons. The question we address in this article is the persistence of a finite-time singularity for the photon distribution when the plasma is no longer described by a Maxwellian distribution, but instead by a linearised Vlasov equation.
  
The photon gas is characterized by an arbitrary initial distribution with averaged energy $E_{ph}$. This may, in particular, be a Planck distribution with a characteristic temperature
$T_{ph}$. We assume that the electron gas is described by a Maxwellian distribution with temperature $T_{e}$. Furthermore the system is assumed to be homogeneous and closed, which means that no particles leave or enter the system.
To take into account the role of Compton scattering in the convergence to equilibrium between photons and non-relativistic electrons, Kompaneets~\cite{KO}
proposed, for a given electron distribution $f_0$,
the following kinetic equation for the photon distribution
function $n_{\ko}(\hbar\omega,t)$ in a unit volume:
\begin{eqnarray}\nonumber
&&\frac{\partial n_{\ko}}{\partial t}=-\int {d^3 p}\int
dW ~ n_{\ko}(\hbar\omega,t)(1+n_{\ko}(\hbar\omega^\prime,t))f_{0}(\varepsilon)\\
&& \qquad+\int {d^3p}\int dW  ~
n_{\ko}(\hbar\omega^\prime,t)(1+n_{\ko}(\hbar\omega,t))f_{0}(\varepsilon+\hbar(\omega-\omega^\prime)),\label{a1}
\end{eqnarray}
with $n_{\ko}(\hbar\omega,t)=
\int {d^3r}~ n_{\ko}(\hbar\omega,{\r},t)$
and the integrand is the photon space-density for a given energy $\hbar\omega$ and time $t$.
Here $f_{0}(\varepsilon)$ is the Maxwellian distribution function for the free electrons with temperature $T_e$ and energy
$\varepsilon=p^2/(2m)$, and $dW$ is the differential probability of transition from a state of energy $\hbar \omega^\prime$ into another state of energy $\hbar \omega$, compatible with the laws of conservation of energy and momentum (for details, see Ref.~\cite{KO}).
 In Eq.~(\ref{a1}), inelastic emission and absorption processes have not been taken into
account, so that the transitions are produced exclusively by Compton scattering processes. Since free electrons do not absorb or emit, but only scatter photons, the total number of photons is conserved, which has to be satisfied by Eq.~(\ref{a1}).
In what follows, it is convenient to use dimensionless variables related to the  electron temperature $T_e$:
$$
\label{newvari}
\t=\frac{\k\T}{mc^2}\frac{c}{l}t=\alpha t,\qquad x=\frac{\hbar\omega}{\k\T},
$$
where $\t$ is referred to as a rescaled time, $x>0$ as the energy of photons which have the distribution function $n_{\ko}(x,\t)$, and $l$ is the Compton range, determined by the total Thomson scattering cross section. So
$l=(\rho\sigma_{\T})^{-1}$, where $\sigma_{\T}=\frac{8\pi}{3} \left(\frac{e^2}{ mc^2}\right)^2$ in Gaussian units, and $\rho$ is the average electron density in configuration space.\\
In terms of these variables, the condition that the total number of photons $N$ is constant can be written in the form~\cite{ZL}
$$
\frac{\partial N}{\partial\t}=0\;\;\mbox{with}\;\; N=\int_0^\infty n_{\ko}(x,\t)x^2 \, dx.
$$
Kompaneets~\cite{KO} considered the non-relativistic case when $ \k\T \ll mc^2$, and he assumed that the energy transferred during each scattering process is small compared to $\hbar\omega$, i.e., $|\Delta|\ll\omega$ where $\Delta = \omega^\prime-\omega$. Expanding the integrand of Eq.~(\ref{a1}) in power series in $\Delta$ up to second order, Eq.~(\ref{a1}) becomes
\begin{eqnarray}\label{a11}
&&\frac{\partial n_{\ko}}{\partial t} = \left(\frac{\partial n_{\ko}}{\partial
x}+n_{\ko}(1+n_{\ko})\right)
 \frac{\hbar}{\k\T}\int {d^3 p}\int dW f_0(\varepsilon)\Delta\\
&&+\left(\frac{{\partial}^2 n_{\ko}}{\partial x^2}+2(1+n_{\ko})\frac{\partial n_{\ko}}{\partial
x}+n_{\ko}(1+n_{\ko})\right )
  \frac{\hbar^2}{2(\k\T)^2}\int {d^3 p}\int dW f_0(\varepsilon){\Delta}^2.\nonumber
 \end{eqnarray}
The integrals in Eq.~(\ref{a11}) can be explicitly computed (see Appendix A). As a consequence,
Kompaneets~\cite{KO} obtained the following equation for the photon distribution function $n_{\ko}(x,\t)$:
\begin{equation}\label{kz1}
  \frac{\partial n_{\ko}}{\partial \t}=\frac{1}{x^2}\frac{\partial}{\partial x}\left[  x^4
  {\left(\frac{\partial n_{\ko}}{\partial x}+n_{\ko}+n_{\ko}^2\right)}\right].
\end{equation}
The terms $n_{\ko}$ and $n_{\ko}^2$ in Eq.~(\ref{kz1}) correspond to the {\it spontaneous scattering} (Compton effect) and the {\it induced scattering}, respectively (see Ref.~\cite{Ibr} and references therein).
It is easy to check that in the stationary case $\partial_{\t} \bar{n}_{\ko}=0$, the solution of Eq.~(\ref{kz1}) is the Planck distribution
$ \bar{n}_{\ko}(\hbar\omega)= 1/(\exp(\hbar\omega/\k\T)-1)$.

In Ref.~\cite{ZL} Zel'dovich and Levich considered the convergence to equilibrium in a system without absorption and emission. Later they were interested in the expression of the photon distribution function in the case of large occupation numbers. They assumed that the occupation numbers are extremely large, so that $n_{\ko}\gg 1$ and $n_{\ko}^2\gg |\partial n_{\ko}/\partial x|$.
These two approximations lead to a drastic simplification of Eq.~(\ref{kz1}):
It
reduces to the inviscid Burgers' equation:
\ba\label{z1}
&&\frac{\partial n_{\ze}}{\partial \t}=\frac{1}{x^2}\frac{\partial}{\partial x}( x^4n_{\ze}^2),\\
&& x^2 n_{\ze}(x,0)=F(x).\nonumber
\ea
Using Eq.~(\ref{z1}), Zel'dovich and Levich~\cite{ZL} evidenced a drastic
increase of the photon distribution function $n_{\ze}$ at $x=0$ in absence of absorption. They refer to this accumulation of photons in the vicinity of the zero energy, as a Bose condensation. This kinetic condensation depends on the form of the initial photon distribution
$F(x)$.
For a specific form of initial distribution,
 a shock-wave profile for $n_{\ze}$ as  the photon energy function occurs. The form of $n_{\ze}$ is extremely non-uniform across the frequency spectrum and is substantially affected by absorption if it is not negligible.
 Using the method of characteristics, Zel'dovich and Levich~\cite{ZL} found the following solution of Eq.~(\ref{z1}):
$$
x=F^{-1}(x^2 n_{\ze})-2\tilde{t} x^2 n_{\ze}.
$$
Considering the Planck distribution
 $n_{\ze}(x,0)=(\exp(\mu x)-1)^{-1}$, where $\mu=T_e/T_{ph}$, for the photons with temperature $T_{ph}$
 and using the approximation
$n_{\ze}(x,0)\approx (\mu x)^{-1}$ for $\mu x \ll 1$, they have inferred the existence of a Bose condensation in the state $x=0$ (see also Refs.~\cite{tsin04,YP,gruz}) for the minimal critical time
\be\label{time}
\t_{*}=\frac{\mu}{2}.
\ee

However, in this special case of the initial Planck distribution the terms, $|\partial n_{\ze}/\partial x|$ and $n_{\ze}^2$ are of the same order in $x$ at $\t=0$, and its ratio is of order $\mu$.
The inequality $n_{\ze}^2\gg\vert\frac{\partial n_{\ze}}{\partial x}\vert$ is valid only for $\t\sim \mu/2$.

For the laboratory plasma this minimal critical time $\t_{*}$ can be extremely long invalidating the assumption for deriving Eq.~(\ref{z1}).
Moreover, in the case of a weakly collisional plasma, the relaxation to a Maxwellian distribution for the electrons can also be extremely long.

In this paper we investigate the Kompaneets equation~(\ref{a1}) by taking into account a perturbation of the Maxwellian distribution by plasma waves. We found an exact solution for the photon distribution function. From this solution we deduce a stability of the solution with respect to perturbation by the plasma waves in the long-wavelength limit.

\section{Solution of Eq.~(\ref{kz1}) in the case of large occupation numbers
$\mathbf{n_{\ko}(x,\t)\gg 1}$}

In general the rescaled critical time  $\t_{*}=\mu/2$ [see Eq.~(\ref{time})] is very large compared with the lifetime of hot and low density laboratory plasmas. For instance, for $T=10^5 {\rm K}$ and $\rho=10^{13}{\rm cm}^{-3}$, this characteristic time might of the order of two days~\cite{CH}. It is worth noting that the critical time varies as $1/(\rho T)$. The hotter and the more dense the plasma is, the lower is the critical time.
Here we are interested in the dynamics corresponding to Eq.~(\ref{kz1}) for $\t\leq \mu/2$. The aim is to find a photon distribution function taking into account the interaction of photons with
electrons via Compton scattering in the case when the Maxwellian distribution function for the electrons is perturbed by a solution of the linearised Vlasov equation in the long-wavelength limit $\lambda\rightarrow\infty$.

Note that the photon temperature is related to the radiation spectrum of the plasma. Radiation in plasmas have various sources, e.g., cyclotron radiation, bremsstrahlung, etc. At high radiation frequency, i.e., at high photon temperature, a dominant mechanism is the recombination process in which electrons undergo a free-bound transition with the positive ions of the plasma.

We consider a special case of the initial Planck distribution in the frequency region, which corresponds to large occupation numbers $n_{\ko}\gg 1$, and we set $n_{\ko}(x,0)=1/(\exp(\mu x )-1)\approx (\mu x)^{-1}$ for $\mu x\ll 1$.
 In this limit, which is known as the case with dominating induced scattering regime~\cite{Ibr},  Eq.~(\ref{kz1}) becomes
\begin{equation}\label{kz1EC}
  \frac{\partial n_{\ko}}{\partial \t}=\frac{1}{x^2}\frac{\partial}{\partial x}\left[  x^4
  {\left(\frac{\partial n_{\ko}}{\partial x}+n_{\ko}^2\right)}\right], \,\, \mbox{ with }\,\,
  n_{\ko}(x,0)=\frac{1}{\mu x}.
\end{equation}
To find a solution of Eq.~(\ref{kz1EC}) we introduce the new variables
$$z=-\ln x,\quad\quad\tau =\t,\quad\quad \mbox{and}\quad\quad \tilde{n}(z,\tau)=x\,n_{\ko}(x(z),\t(\tau)).$$
Then one gets for $\nt(z,\tau)$ the following nonlinear equation  with constant coefficients:
\begin{equation}\label{ntilde}
\frac{\partial\nt}{\partial\tau}=\frac{\partial^2\nt}{\partial z^2}-2\nt\frac{\partial\nt}{\partial z}+2{\nt}^2-\frac{\partial\nt}{\partial z}-2\nt, \,\,\mbox{ with }\,\, \nt(z,0)= 1/\mu.
\end{equation}

We are looking for a solution of Eq.~(\ref{ntilde}) in the form:
\begin{equation}\label{ntilde1}
\nt(z,\tau)=\varphi(\tau)+\psi(\tau)\exp(z).
\end{equation}
Inserting this form of solution~(\ref{ntilde1}) into Eq.~(\ref{ntilde}), we obtain
\begin{equation}\label{ntilde2}
(\varphi^{\prime}_{\tau}-2{\varphi}^2+2\varphi)=
(-{\psi}^{\prime}_{\tau}+2\varphi\psi-2\psi)\exp(z).
\end{equation}
The left-hand side of Eq.~(\ref{ntilde2}) is a function of $\tau$, contrary to the right-hand side, which is a function of two variables $\tau$ and $z$. Therefore the left- and the right-hand sides of Eq.~(\ref{ntilde2}) must be equal to zero. Consequently, we get two ordinary differential equations with initial conditions:
\begin{eqnarray}\label{ntilde3}
&&\varphi^{\prime}_{\tau}-2{\varphi}^2+2\varphi=0,\qquad 0\leq\tau < \mu/2,\nonumber\\
&&\varphi(0)=1/\mu,
\end{eqnarray}
and
\begin{eqnarray}\label{ntilde4}
&&{\psi}^{\prime}_{\tau}-2\varphi\psi+2\psi=0,\qquad 0\leq\tau < \mu/2,\nonumber\\
&&\psi(0)=0.
\end{eqnarray}
The solutions of Eq.~(\ref{ntilde3}) and Eq.~(\ref{ntilde4}) are correspondingly
$$
\varphi(\tau)=\frac{1}{1-(1-\mu)\exp(2\tau)},\qquad\psi(\tau)=0.$$
Consequently,  the solution of Eq.~(\ref{kz1EC}) is
\begin{equation}\label{ntilde5}
n_{\ko}(x,\t)=\frac{1}{x}\frac{1}{1-(1-\mu)\exp(2\t)},
\end{equation}
which coincides with the invariant solution (28) in Ref.~\cite{Ibr}.
This solution is valid for $0<\mu x\ll 1$, which implies that $n_{\ko}\gg 1$.
Here we notice that $\partial n_{\ko} /\partial x$ is of the same order as $n_{\ko}^2$. When $\mu>1$ this solution does not have any singularity, whereas when $\mu<1$, i.e., when $T_{ph}>T_{e}$, there is a finite-time singularity at $\t_{\tiny{s}}=-\ln\sqrt{1-\mu}$. This time is always larger than the minimal critical time $\t_{*}=\mu/2$~\cite{ZL}.
When $\mu=1$ one obtains the stationary solution of Eq.~(\ref{kz1EC}).

\section{Photon distribution extended to the case of Vlasov plasma wave oscillations}

In the previous section, it was assumed that the distribution of the electrons $f_0$ was Maxwellian. However if the collisionality is small, the relaxation to a Maxwellian distribution is extremely long. For collisionless plasmas, the evolution of the distribution is determined by the Vlasov equation.
Here we restrict ourselves to the solutions of the linearised Vlasov equation in the
long-wavelength (low-frequency) approximation for the plasma oscillations.
Then the zeroth order approximation coincides with the space-homogeneous Maxwellian distribution $f_0(\v)$ and we consider a perturbation of this distribution. To this aim we denote by
\begin{equation}\label{vlasove}
f(\r,\v,\t)   =  f_0(\v)  +  \sigma f^*(\r,\v,\t),
\end{equation}
the electron distribution function, where $f^*(\r,\v,\t)$ is a solution of the  linearised Vlasov equation.
Similarly the photon distribution function is given by
\begin{equation}\label{vlasovph}
n(x,\t) =   n_{\ko}(x,\t)    +  \sigma n^*(x,\t) .
\end{equation}
Here $n_{\ko}(x,\t)$ is a non-stationary solution of Eq.~(\ref{kz1EC}) given by Eq.~(\ref{ntilde5}).
We notice that in the long-wave limit the effective collisions due to photon-plasmon interactions are attenuated and we neglect them (see Refs.~\cite{Vl,L2} for more detail).

In a nutshell, our strategy is to obtain an equation for the perturbation $n^*(x,\t)$ by inserting the expressions~(\ref{vlasove}) and (\ref{vlasovph}) into Eq.~(\ref{a1}). Then our goal is to solve this equation in order to estimate the impact of plasma wave oscillations on the background non-stationary photon distribution
$n_{\ko}(x,\t)$.

The linearised Vlasov-Poisson equation has the form~\cite{AAVl,Vl}
\begin{eqnarray}\label{a5}
\frac{\partial f^*}{\partial t}&+&{\v}\cdot \frac{\partial
f^*}{\partial{\r}}-\frac{e{\mathbf{E^*}}}{m}\cdot\frac{\partial f_0}{\partial{\v}}=0,\\
 \nabla \cdot {\mathbf{E^*}}  &=& -4\pi e\rho \int f^*({\r,\v,}{t})d^3v, \nonumber
\end{eqnarray}
where
$$\label{a6bis}
f_0({\v})=\left(\frac{m}{2\pi \k\T}\right)^{3/2}\exp\left(-\frac{m{\v}^2}{2\k\T}\right),
$$
and $\rho$ is the average electron density.
We consider plasma oscillations propagating in the $r_1$-direction, i.e.,
\begin{eqnarray}\label{a6}
f^*({\r,\v,}{t}) &=& f_{k\omega}({\v})\exp(-i\omega t+i k r_1),\\
E^*_{r_1}({\r},t) &=& E_{k\omega} \exp(-i\omega t+i k r_1),\nonumber \\
E_{r_2} &=& E_{r_3}=0.\nonumber
\end{eqnarray}
In particular we are interested in the long-wavelength limit ($k\rightarrow 0)$. Following Kvasnikov~\cite{Vl} we insert~ Eq.~(\ref{a6}) into Eq.~(\ref{a5}) to obtain  the expressions for the amplitudes $f_{k\omega}$:
\begin{eqnarray}
\label{App23-bis}
&& -i\omega f_{k\omega}+i k v_1 f_{k\omega}-\frac em\frac{\partial f_0}{\partial v_{1}}E_{k\omega}=-\varepsilon f_{k\omega}
\mid_{\varepsilon\rightarrow 0},\\
&& ikE_{k\omega}=-4\pi e\rho\int f_{k\omega}(\v)d^3 v. \label{App23-bis1}
\end{eqnarray}
Then from Eq.~(\ref{App23-bis})  follows :
\begin{equation}
\label{f0}
f_{k\omega}=i\frac{\frac{e}{m}\frac{\partial f_0}{\partial v_{1}}}{\omega+i\varepsilon-v_{1}k}E_{k\omega}.
\end{equation}
Inserting Eq.~(\ref{f0}) into Eq.~(\ref{App23-bis1}), we obtain
$$
\label{Eko}
ikE_{k\omega}=-4\pi ei\frac{e\rho}{m}E_{k\omega}\int\frac{\frac{\partial f_0}{\partial v_{r_1}}}{\omega-v_{r_1}k+i\varepsilon}d^3 v.
$$
Excluding the trivial solution $f_{k\omega}=E_{k\omega}=0$, which corresponds to the unperturbed case, one gets the dispersion relation $\omega=\omega({k})$.
After integration with respect to $v_{r_2}$ and  $v_{r_3}$, we get an equation for $\omega=\omega({k})$ in the following form:
\be\label{App25}
1-\frac{\omega_0^2}{k}\frac {m}{\k\T}\left(\frac{m}{2\pi\k\T}\right)^{\frac 12}
\int_{-\infty}^\infty dv_{r_1}\frac{v_{r_1}}{\omega-v_{r_1}k+i\varepsilon}
\exp\left(-\frac{m v_{r_1}^2}{2\k\T}\right)=0,
\ee
where $\omega^2_0=4\pi e^2\rho/m$.
Calculating the integral in Eq.~(\ref{App25}) we get:
\be\label{App28}
1-\frac{\omega_0^2}{\omega^2}\left(1+\frac{3\k\T}{m\omega^2}k^2+\frac{m}{\k\T\omega^4} k^4\overline{\left(\frac{v_{r_1}^6}{1-\frac{k}{\omega}v_{r_1}}\right)}\right)+i J(k,\omega)=0,
\ee
where $J(k,\omega)$ is
$$\label{Iko}
J(k,\omega)=\sqrt{\frac{\pi}{2}}\frac{\omega_0^2 m^{3/2}}{k^3(\k\T)^{3/2}} \omega\exp\left
(-\frac{m \omega^2}{2k^2\k\T}\right),
$$
and the bar in Eq.~(\ref{App28}) denotes the mean value with respect to the Maxwellian distribution.

In the long-wavelength approximation, when $k^2\ll \frac{m}{\k\T}\omega^2$, we neglect the term proportional to $k^4$ in Eq.~(\ref{App28}). Since $J(k,\omega)\neq 0$, Eq.~(\ref{App28})
has no real solution. We consider the solutions of Eq.~(\ref{App28}) in the form $\omega=\Omega-i\gamma$ where $\gamma/\Omega\ll 1$. Then
\be\label{App29}
\frac{1}{\omega^2}=\frac{1}{(\Omega-i\gamma)^2}\approx \frac{1}{\Omega^2}+i\frac{1}{\Omega^2} \frac{2\gamma}{\Omega},
\ee
From Eq.~(\ref{App29}) it follows that
\be\label{App30}
1-\frac{\omega_0^2}{\Omega^2}\left(1+\frac{3\k\T}{m\Omega^2}k^2\right)-i\frac{\omega_0^2}{\Omega^2}+i J(k,\Omega)=0.
\ee
Considering the real and imaginary parts of Eq.~(\ref{App30}) we finally obtain
\ba\label{App31}
\Omega(k)&=&\omega_0\left(1+\frac{3\k\T}{2m\omega_0^2}{k^2}+\cdots\right), \\
\gamma(k)&\approx &\frac{\omega_0}{2}J(k,\omega_0)=\frac12\sqrt{\frac\pi 2}\frac{\omega_0^4}{k^3}\frac{m^{3/2}}{(\k\T)^{3/2}}\exp\left(-\frac{m\omega_0^2}
{2k^2\k\T}\right),
\ea
where $\gamma$ is the Landau damping, which vanishes in the long-wavelength limit ($k\rightarrow 0$).
We notice that for $k\neq 0$ the problem is much more complicated (see Refs.~\cite{L1,L2}).
From Eq.~(\ref{App31}) it follows that the first approximation for $\omega(k)$ is $\omega(k)=\omega_0$.

Without loss of generality we assume that $E_{0\omega}=E$.
Taking into account Eq.~(\ref{f0}) we get for $f_{0\omega}$
in the long-wavelength limit ($k\rightarrow 0$)
$$\label{a7}
 f_{0\omega}=i\frac{eE}{m\omega_0}\frac{\partial f_0}{\partial v_{1}}.
$$
In summary, we consider the following solution of the linearised Vlasov equation~(\ref{a5}):
\be\label{vlasovend}
 f^*(\r,{\v,}{\t})=-\frac{eE}{\omega_0\k\T}\sin(\omega_0\alpha^{-1}\t)v_{1} f_0({\v})=\theta(\tilde{t})v_1f_0(\v),
\ee
where $\omega_0=\sqrt{4\pi e^2\rho/m}$ is the Langmuir frequency.

Next we insert $n(x,{\t})$ given by Eq.~(\ref{vlasovph}) and $f(\r,\v,t)$ given by Eq.~(\ref{vlasove}) with $f^*(\r,{\v,}{\t})$ obtained in Eq.~(\ref{vlasovend}) in  Eq.~(\ref{a1}), to get an equation for the unknown function ${n}^*(x,{\t})$ :
\begin{eqnarray}\label{b2}
 \frac{\partial n^*}{\partial \t}&=&\alpha^{-1}\left[\gamma_2\frac{{\partial}^2n^*}{\partial
 x^2}+\left(\gamma_1+2\gamma_2(n_{\ko}+1)\right)\frac{\partial n^*}{\partial x}\right.\nonumber\\
 &&\left. +\left(\gamma_1(1+2n_{\ko})\!+\!\gamma_2\left(1+2n_{\ko}+2\frac{\partial n_{\ko}}{\partial x}\right)\right)n^*\!+\!\Phi(x,\t)\right],
\end{eqnarray}
where (see Appendix A):
\begin{eqnarray*}
\gamma_1&=&\frac{\hbar}{\k\T}\int d^3 v \int dW f_0({\v})\Delta=\alpha x(4-x),\\
\gamma_2&=&\frac{1}{2}\left(\frac{\hbar}{\k\T}\right)^2\int d^3v \int dW f_0({\v}){\Delta}^2=\alpha x^2,
\end{eqnarray*}
\begin{eqnarray}\label{korr1}
\alpha^{-1}\Phi(x,\t)&=&
\left(\!\frac{\partial n_{\ko}}{\partial x}+n_{\ko}(1+n_{\ko})\right)\frac{\hbar\alpha^{-1}}{\k\T}\int\! d^3 p\int\! dW\theta(\t)v_{1}f_0({\v})\Delta, \nonumber\\
&=&
\left(\frac{\partial n_{\ko}}{\partial x}+n_{\ko}(1+n_{\ko})\right)\beta x\sin(\omega_0\alpha^{-1}\t).
\end{eqnarray}
Here $\beta=(eE c)/(2\omega_0\k\T)$.
Plugging in Eq.~(\ref{b2}) the expressions for $\gamma_1$, $\gamma_2$ and $\Phi$, and keeping only the dominant terms when $\mu x\ll 1$ and $\t<\mu/2$, one gets for $n^*(x,\t)$
 the following equation:
\ba\label{initial1}
\frac{\partial n^*}{\partial \t}&=&x^2\frac{\partial n^*}{\partial x^2}+x\left(4+\frac{2a}{1-2(a-1)\t}\right)\frac{\partial n^*}{\partial x}+\frac{6a}{1-2(a-1)\t}\,n^*\nonumber\\
&&+\frac{a(a-1)\beta}{x}\,\frac{1+2\t}{[1-2(a-1)\t]^2}\sin(\omega_0\alpha^{-1}\tilde{t}), \nonumber
\ea
where $a=1/\mu$ and with $n^*(x,0)=0$.
The  changes of variables, valid for $1-2(a-1)\t>0$,
\begin{eqnarray}\label{chenge1}
&& \eta=\tilde{t},\nonumber \\
&& \label{change2}
\xi=-\ln x+\frac{a}{a-1}\ln[1-2(a-1)\t],\nonumber \\
&& \label{change3}
 u(\xi,\eta)=n^*(\xi,\eta)[1-2(a-1)\eta]^{3a/(a-1)},
\end{eqnarray}
lead to equation for $u(\xi,\eta)$ with constant
coefficients:
\ba\label{initial2}
&&\frac{\partial u}{\partial\eta}=\frac{\partial^2 u}{\partial \xi^2}-3\frac{\partial u}{\partial\xi}+
a(a-1)\beta\exp(\xi)
\frac{1+2\eta}{[1-2(a-1)\eta]^\frac{2}{1-a}}
\sin(\omega_0\alpha^{-1}\eta),\nonumber\\
&&{u}^*(\xi,0)=0,\,\,\,-\infty<\xi<\infty,\,\,\,\eta>0. \nonumber
\ea
The second change of variables is a well-known trick to eliminate the first derivative:
$u(\xi,\eta)=\exp(\frac32\xi-\frac94\eta) u^{*}(\xi,\eta)$.  The equation for $u^{*}(\xi,\eta)$ is then
\ba\label{initial3}
&&\frac{\partial {u}^*}{\partial \eta}=\frac{\partial^2 u^{*}}{\partial \xi^2}+
a(a-1)\beta\exp\left(-\frac{\xi}{2}+\frac{9\eta}{4}\right)
\frac{1+2\eta}{[1-2(a-1)\eta]^\frac{2}{1-a}}
\sin(\omega_0\alpha^{-1}\eta),\nonumber\\
&&{u}^*(\xi,0)=0,\,\,\,-\infty<\xi<\infty,\,\,\,\eta>0.
\ea
Using the Fourier transformation one  finds the solution of the non-homogeneous linear partial differential equation~(\ref{initial3}) with constant coefficients:
\ba\label{initial4}
{u}^*(\xi,\eta)&=&
\frac{a(a-1)\beta}{2\sqrt{\pi}}\!\!\int_0^{\eta}d{\eta}^{\prime}\!\!\int_{-\infty}^{\infty}d{\xi}^\prime
\sin(\omega_0\alpha^{-1}{\eta}^\prime)\frac{1+2{\eta}^\prime}{[1-2(a-1)\eta']^{\frac{2}{1-a}}}\nonumber \\
&& \qquad \qquad \qquad \qquad \qquad \qquad \times \frac{\exp\left(-\frac{(\xi-{\xi}^\prime)^2}
{4(\eta-{\eta}^\prime)}\!\!+\!\!\frac{9{\eta}^\prime}{4}-\frac{{\xi}^\prime}{2}\right)}
{\sqrt{\eta-{\eta}^\prime}}\nonumber\\
&=& a(a-1)\beta\exp\left(-\frac{\xi}{2}+\frac{\eta}{4}\right)I(\eta,a), \nonumber
\ea
where
$$
I(\eta,a)=\int_0^{\eta}
d{\eta}^{\prime} \exp(2\eta') \sin(\omega_0\alpha^{-1}{\eta}^\prime)(1+2{\eta}^\prime)[1-2(a-1)\eta']^{\frac{2}{a-1}}.
$$
Finally, from Eq.~(\ref{change3}) it  follows the expression for the photon distribution
function $n^{*}(x,\t)$, which takes into account the effect of the Vlasov plasma wave oscillations:
\be\label{rezult}
n^*(x,\t)=\frac{a}{x}(a-1)\beta\frac{{\rm e}^{-2\t}}{[1-2(a-1)\t]^{\frac{2a}{a-1}}}I(\t,a).
\ee
Then Eq.~(\ref{vlasovph}) leads to
\ba\label{initial5}
n(x,{\t})&=&n_{\ko}(x,\t)+\sigma {n}^*(x,{\t}),\nonumber\\
&=&\frac{a}{x}\frac{1}{1-2(a-1)\t}\left[1+\sigma (a-1)\beta
\frac{{\rm e}^{-2\tilde{t}}}{[1-2(a-1)\t]^{\frac{a+1}{a-1}}}I(\t,a) \right].
\ea
The integral $I(\t,a)$ in Eq.~(\ref{initial5}) can be computed using the incomplete Gamma function~\cite{nielson}. However the explicit expression of these integrals are not very useful for the analysis. In what follows, we discuss the two main cases $\mu >1$ and $\mu <1$ using a combination of numerical computation and series expansions of the integrals.

\begin{itemize}
\item If $\mu<1$, i.e., $T_{ph}>T_e$, the Bose condensation is amplified by the perturbation of the Maxwellian distribution since the exponent $(a+1)/(a-1)$ is positive. Therefore, the solution for $n_{\ko}(x,\t)$ is unstable. On Fig.~\ref{fig1}, we represent the computation of $n^*(x,\tilde{t})$ given by Eq.~(\ref{rezult}) for $\mu=1/2$ and $\omega_0\alpha^{-1}=100$. It is clear that there is a divergence of the perturbed number of photons when $\tilde{t}$ approaches $1/2$. Note that this threshold is different from the one in Eq.~(\ref{time}). For $\mu=1/2$, a series expansion of the integral given by Eq.~(\ref{initial5}) can be obtained when $\omega_0\alpha^{-1}$ is large:
$$
I(\t,2)=\frac{1}{\omega_0\alpha^{-1}}-\frac{\cos( \omega_0\alpha^{-1}\t)}{\omega_0\alpha^{-1}}{\rm e}^{2\t}(1-2\t)^2(1+2\t)+O\left(\frac{1}{(\omega_0\alpha^{-1})^2}\right),
$$
which leads to an approximation of $n^*(x,\tilde{t})$:
$$
n^*(x,\tilde{t})=\frac{2\beta}{x}\frac{1}{\omega_0\alpha^{-1}}\left( \frac{{\rm e}^{-2\t}}{(1-2\tilde{t})^4}-\frac{(1+2\tilde{t})}{(1-2\tilde{t})^2}\cos (\omega_0\alpha^{-1}\tilde{t})\right)+O\left(\frac{1}{(\omega_0\alpha^{-1})^2}\right).
$$
It should be noted that this expression is a very good approximation of $n^*(x,\t)$ in a wide range of values of $\omega_0\alpha^{-1}$.
From this expression, we see that initially, the dominant contribution is caused by the plasma wave oscillations. In this regime, we argue that
from the measurement of the photon distribution $n(x,\t)$, one can have access to the plasma density variations. Near the singularity, the dominant term is proportional to $(1-2\t)^{-4}$, and this divergence of $n^*(x,\t)$ is much faster than the unperturbed distribution $n_{\ko}(x,\t)$.
\begin{figure}
\centering
\includegraphics[height=0.37\textheight]{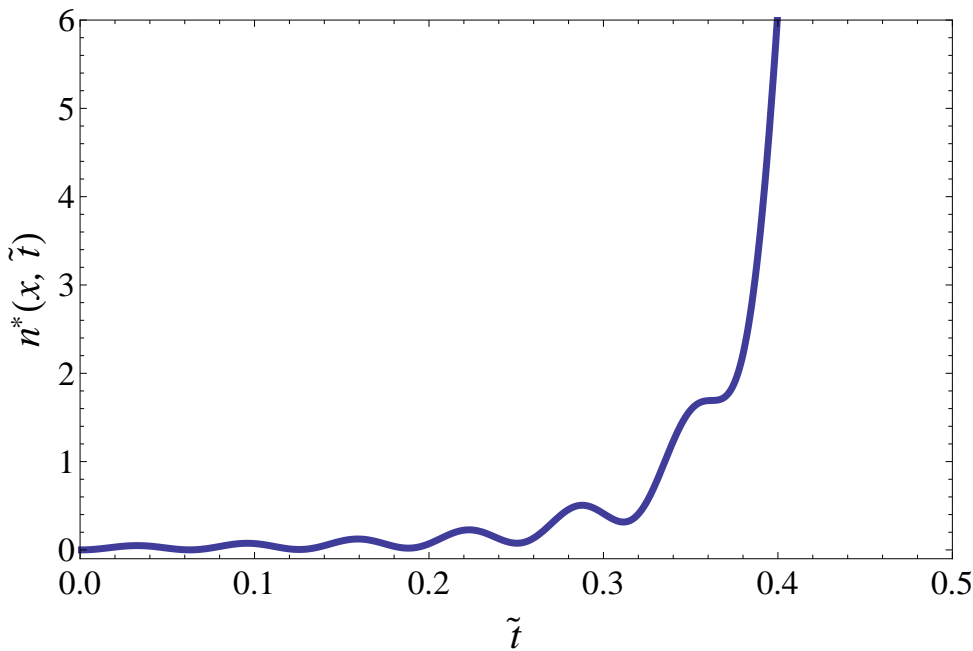}
\caption{$n^*(x,\tilde{t})$ for $x=\beta$ given by Eq.~(\ref{rezult}) as a function of $\tilde{t}$ for $\mu=1/2$ and $\omega_0 \alpha^{-1}=100$.}
\label{fig1}
\end{figure}
\item For $\mu>1$, which corresponds to the case $T_e>T_{ph}$, the resulting perturbation does not have any singularity. So the solution for the photon distribution is rather stable with respect to perturbations of the Maxwellian distribution for the electrons (see Fig.~\ref{fig2}). It should be noted that the values of $\sigma$ are restricted to the case, when  $\sigma n^*$ is much smaller than $n_{\ko}$. So $\sigma$ has be chosen such that $\sigma \ll 1/[(1-a)\beta \mbox{max }_{\t\in [0,1/2]}[[1-2(a-1)\t]^{(1+a)/(1-a)} \vert I(\tilde{t},a)\vert]$.
\begin{figure}
\includegraphics[height=0.37\textheight]{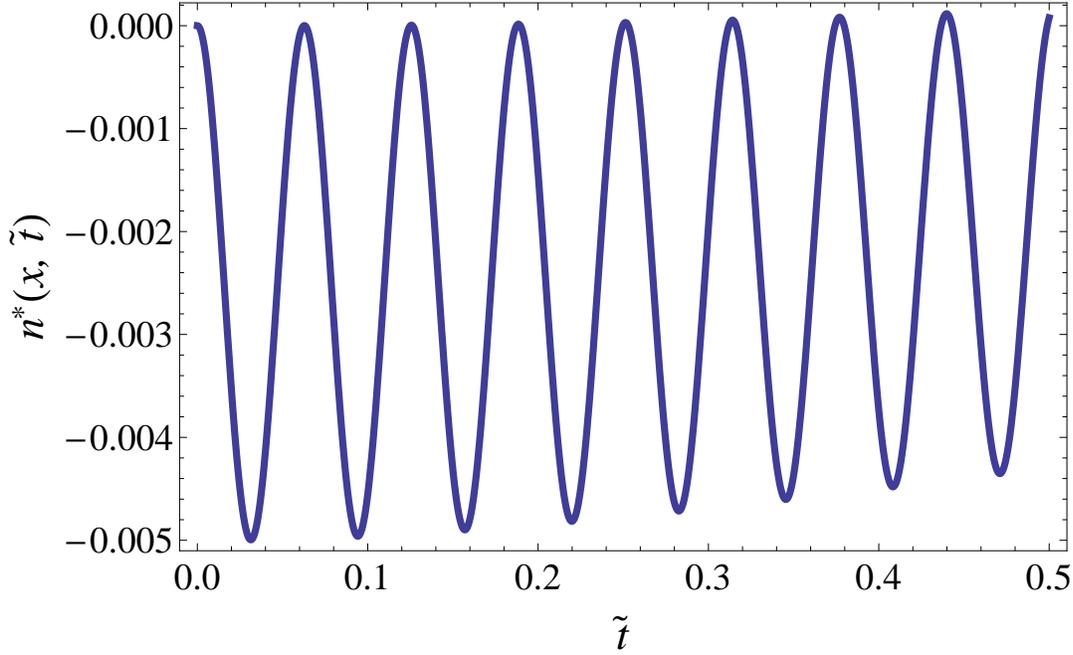}
\caption{$n^*(x,\tilde{t})$ for $x=\beta$ given by Eq.~(\ref{rezult}) as a function of $\tilde{t}$ for $\mu=2$ and $\omega_0\alpha^{-1}=100$.}
\label{fig2}
\end{figure}
\item If $\mu=1$, i.e., $T_e=T_{ph}$, the Vlasov plasma wave oscillations have no effect on the distribution of photons. It corresponds to a stationary case.
\end{itemize}

\section*{Conclusion}

We considered some modifications of the Kompaneets  equation for the photon distribution function when plasma wave oscillations perturb the distribution of the electrons.
We have shown that the solution of the modified Kompaneets equation in the limit of large occupation numbers $n(x,\t)\gg 1$ can be unstable for sufficiently large times around the zero energy $x=0$. In the unstable case, Bose condensation occurs. This instability is generated by the diffusion term in the equation for the photon distribution.

\appendix

\section{Computation of the integrals of Eq.~(\ref{a11})}

Following Kompaneets~\cite{KO} we calculate the coefficients
\ba\label{g1}
\gamma_1&=&\frac{\hbar}{\k\T}\int d^3 v \int dW f_0({\v})\Delta,\nonumber \\
\gamma_2&=&\frac{1}{2}\left(\frac{\hbar}{\k\T}\right)^2\int d^3v \int dW f_0({\v}){\Delta}^2. \nonumber
\ea
The laws of momentum and energy conservation in the non-relativistic case are written as:
\begin{eqnarray*}\label{A1}
\frac{\h}{c}\n+\p&=&\frac{\hbar\omega^\prime}{c}\n^\prime+\p^\prime,\nonumber\\
\h+\frac{p^2}{2m}&=&\hbar\omega^\prime+\frac{{p^\prime}^2}{2m},
\end{eqnarray*}
where $\varepsilon=\frac{p^2}{2m}$, and $\p$ and $\p^\prime$ are the momenta of the electron before and after the collision, $\n$ and $\n^\prime$ are unit vectors in the propagation direction of the photon. Eliminating $\p^\prime$ from these equations, we obtain an equation which determines $\omega^\prime$ as a function of $\omega$, $\p$ and the scattering angles. Taking only into account the linear order in $\delta$ one gets
\be\label{A2}
\hbar\delta=-\frac{\hbar c\omega\p\cdot(\n-\n^\prime)+(\h)^2(1-\n\cdot\n^\prime)}{mc^2[1+\frac{\h}{mc^2}(1-\n\cdot\n^\prime)-
\frac{\p{\cdot}{\n}^\prime}{mc}]}.
\ee
Taking into account the Kompaneets approximation $ kT\ll mc^2$, we estimate the terms in the denominator of Eq.~(\ref{A2}).
Since $\hbar\omega\sim 10^4$ eV and $mc^2=0.5\times 10^6$ eV, one gets $\hbar\omega/mc^2\ll 1$.
Consequently,  $\hbar\omega(1-\n\cdot{\n}^\prime)/(mc^2)<2\hbar\omega/(mc^2)\ll 1$ and
$|\p\cdot {\n}^\prime/(mc)|<pc/(mc^2)=c\sqrt{2mE_{\rm kin}}/(mc^2)\simeq \sqrt{3\k T/(mc^2)}\ll 1$.
The first term in the numerator is of order $ (\k\T/mc^2)^{3/2}$ whereas the second term is of order $(\k\T/mc^2)^2$. Then from Eq.~(\ref{A2}) it follows that
 \be\label{A3}
\hbar\delta=-\frac{\hbar\omega}{mc}\p\cdot(\n-\n^\prime).
\ee
Now we calculate $\gamma_2$ using spherical coordinates:
\begin{eqnarray}\label{g22}
&&\gamma_2=\frac{1}{2}\left(\frac{\hbar}{\k\T}\right)^2\int d^3v \int dW f_0({\v}){\Delta}^2,\\
&&=\frac{\pi}{(\k\T)^2}\left(\frac{\h}{mc}\right)^2\int_0^\infty dpp^4\frac{1}{(2\pi m\k\T)^{3/2}}\exp {\left(-\frac{p^2}{2m\k\T}\right)} \nonumber\\
&& \qquad \qquad \times \int_0^\pi d\theta\cos^2\theta\sin\theta \int_\Omega dW\mid \n-\n^\prime\mid^2.\nonumber
\end{eqnarray}
We replace the Compton cross section in the non relativistic approximation by the Thomson cross section
 which is symmetric relative to scattering over the angles $\theta$ and $\pi-\theta$; so that $\int_\Omega dW\n\cdot\n^\prime=0$.
 Here $dW$ is the differential transition probability, compatible with the energy and momentum conservation laws.
 The Thomson cross section does not depend on the energy of the photons. Consequently
 \be\label{g2}
 \gamma_2=\left(\frac{\h}{\k\T}\right)^2\frac{\k\T}{mc^2}\frac{c}{l}=\alpha x^2,
 \ee
where $l$ is the Compton range, which is determined by the total cross section $\frac{8\pi}{3}\left(\frac{e^2}{4\pi\varepsilon_0 mc^2}\right)^2$.
For the expression for $\gamma_1$, we rewrite the kinetic equation for the photons in a conservative form:
\be\label{eqg1}
\left(\frac{\partial n}{\partial t}\right)_C=-\frac{1}{x^2}\frac{\partial(x^2J(x,t))}{\partial x}
\ee
where $J(x,t)$ is the ``flow'' of photons in frequency space. It is easy to show that in the state of total equilibrium, i.e., when  $\left(\frac{\partial n}{\partial t}\right)_C=0$, the solution of Eq.~(\ref{a1}) is $\bar{n}=(\exp(x)-1)^{-1}$, and satisfies $\partial \bar{n}/\partial x=-\bar{n}(1+\bar{n})$.
Let us find $J(x,t)$ in following form: $$J(x,t)=g(x)[\partial n/\partial x+n(n+1)],$$ where the function $g(x)$
must be determined and $n(x,t)$ is the solution of Eq.~(\ref{a11}). Let $g(x)=-x^2$. Substituting Eq.~(\ref{eqg1}) in Eq.~(\ref{a11}) leads to:
\begin{eqnarray*}
\left(\frac{\partial n}{\partial\t}\right)_C&=&\alpha\frac{1}{x^2}\frac{\partial}{\partial x}
\left[x^4\left(\frac{\partial n}{\partial x} +n(n+1)\right)\right]\\
&=&\left(\frac{\partial n}{\partial x} +n(n+1)\right)\alpha x(4-x)+\left(\frac{\partial^2 n}{\partial x^2} +
2\frac{\partial n}{\partial x}(n+1)+n(1+n)\right)\alpha x^2,
\end{eqnarray*}
which resembles Eq.~(\ref{a11}) with the right coefficient $\gamma_2$. The identification leads to $\gamma_1=\alpha x(4-x)$.


\ack
We thanks two referees for their useful remarks and suggestions. GE would like to thank E.A. Dynin for fruitful discussions.

\end{document}